\documentclass[3p,times,procedia]{elsarticle}

\usepackage{nupha_ecrc}
\usepackage{amsmath}
\usepackage{wrapfig}
\volume{00}

\firstpage{1}

\journalname{Nuclear Physics A}

\runauth{J.\ Scott Moreland et al.}

\jid{nupha}

\jnltitlelogo{Nuclear Physics A}

\usepackage{amssymb}
\usepackage[figuresright]{rotating}

\newcommand{\trento}{T\raisebox{-0.3ex}{R}ENTo}
\newcommand{\sig}{\sigma^\mathrm{inel}_{pp}}
\newcommand{\T}{\tilde{T}}
\newcommand{\e}{\varepsilon}

\newcommand{\vn}{\sqrt{\langle v_n^2 \rangle}}
\newcommand{\en}{\sqrt{\langle \varepsilon_n^2 \rangle}}

\usepackage{graphicx}

\begin{document}

\begin{frontmatter}



%% Instructions from Editor: Please use the following \dochead only in the preprint version (e-print arXiv etc.); 
%% use empty \dochead{} when submitting to Nuclear Physics A!
\dochead{XXVIth International Conference on Ultrarelativistic Nucleus-Nucleus Collisions\\ (Quark Matter 2017)}
%%\dochead{}
%% e.g. \dochead{17th International Conference on Dynamical Processes in Excited States of Solids}

\title{Flow in small and large quark-gluon plasma droplets:\\the role of nucleon substructure}

\author{J.\ Scott Moreland, Jonah E.\ Bernhard, Weiyao Ke, and Steffen A.\ Bass}
\address{Department of Physics, Duke University, Durham, NC 27708-0305}

\begin{abstract}
We study the effects of nucleon substructure on bulk observables in proton-lead collisions at the LHC using Bayesian methodology. Substructure is added to the \trento\ parametric initial condition model using Gaussian nucleons with a variable number of Gaussian partons. We vary the number and width of these partons while recovering the desired inelastic proton-proton cross section and ensemble averaged proton density. We then run the model through a large number of minimum bias hydrodynamic simulations and measure the response of final particle production and azimuthal particle correlations to initial state properties. Once these response functions are determined, we calibrate free parameters of the model using established Bayesian methodology. We comment on the implied viability of the partonic model for describing hydrodynamic behavior in small systems.
\end{abstract}


\end{frontmatter}

% \linenumbers

\section{Introduction}

Recent measurements of azimuthal particle correlations in small collision systems show striking similarities to flow signatures observed in gold-gold and lead-lead collisions, leading many to question if the origin of small system correlations is hydrodynamic in nature.
A key difficulty in assessing the model-to-data consistency of hydrodynamic models in light-light and light-heavy collisions is theoretical uncertainty in the QGP initial conditions.
In this work, we use established Bayesian methodology \cite{Higdon:2008cmc} to parametrize and constrain the QGP initial conditions in small collision systems in order to infer initial state properties with reduced bias.

\begin{figure}
  \centering
  \includegraphics{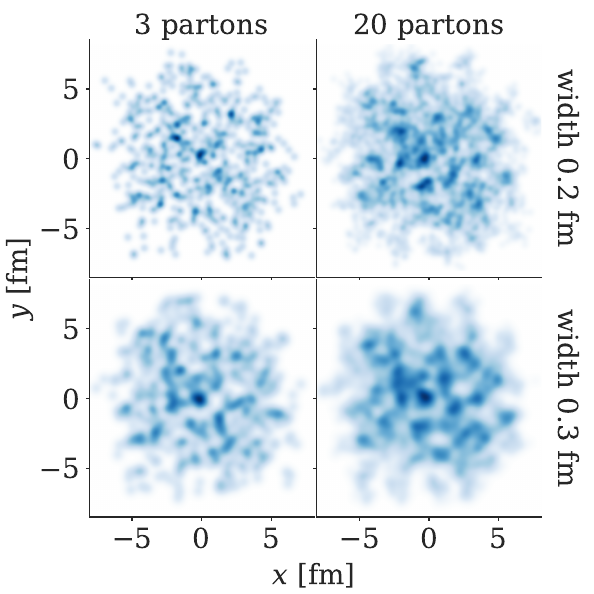} 
  \hspace{8ex}
  \includegraphics{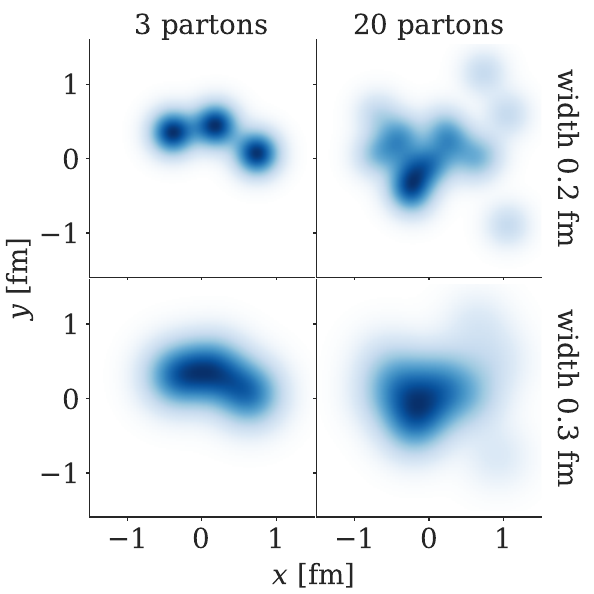}
  \caption{\label{fig:substructure}
    Nuclear thickness functions $T$ with nucleon substructure for a single lead nucleus (left panel) and proton (right panel) for parton widths $v= 0.2, 0.3$~fm (rows) and parton numbers $m=3,20$ (columns).
  }
\end{figure}

The \trento\ initial condition model \cite{Moreland:2014oya} used in this work parametrizes local nuclear entropy deposition at midrapidity according to a reduced thickness function
\begin{equation}
  \label{gen_mean}
  \frac{dS}{d^2x\, \tau_0\, d\eta} \bigg\vert_{\eta=0} \propto\, \left(\frac{\T_A^p + \T_B^p}{2}\right)^{1/p}.
\end{equation}
Here $\T$ is the modified participant thickness function $\T(\mathbf{x}) \equiv \sum_{i=1}^{N_\mathrm{part}} \gamma_i\, T_p(\mathbf{x} - \mathbf{x}_i)$, where $\gamma_i$ is a random weight factor sampled from a Gamma distribution with unit mean and fluctuation standard deviation $\sigma_\mathrm{fluct}$, while $T_p$ denotes the proton thickness function described by a normalized Gaussian distribution with nucleon width $w$. 
The exponent $p$ in Eq.~\eqref{gen_mean} is a tunable entropy deposition factor which takes continuous values $p \in (-\infty, \infty)$ and allows the model to mimic various theory calculations \cite{Bernhard:2016tnd}. 

Participant nucleons in the colliding nuclei are first determined according to the pairwise collision probability
\begin{equation}
  \label{collision_criteria}
  P_\mathrm{coll}(b) = 1 - \exp[-\sigma_{gg} T_{pp}(b)],
\end{equation}
where $T_{pp}$ denotes the proton-proton overlap function, and the cross section parameter $\sigma_{gg}$ is determined to satisfy the experimentally measured proton-proton cross section $\sig = \int d^2b\, P_\mathrm{coll}(b)$.
The participant nucleons in each nucleus are then used to construct the participant thickness functions $\T_A$ and $\T_B$ which are subsequently passed through the generalized mean in Eq.~\eqref{gen_mean} to furnish the initial transverse entropy density.

\section{Nucleon substructure}

The generalized mean parameter $p=0$, which optimally describes charged particle yields and flows in heavy nuclei \cite{Bernhard:2016tnd}, predicts perfectly Gaussian QGP entropy profiles in high energy proton-proton collisions when the interacting protons are modeled as Gaussians. 
By construction, such profiles will never drive anisotropic transverse flow and hence could not be used to investigate hydrodynamic behavior in small collision systems where the system size approaches the proton length scale.

One possible solution to resolve the apparent conflict is to replace Gaussian protons with deformed or lumpy protons \cite{Schenke:2014zha}.
In this work, we extend the \trento\ formalism and replace each nucleon with a fixed number of Gaussian partons.
The new proton thickness function $T_p$ becomes
\begin{equation}
  T_p(\mathbf{x}) = \gamma_i \sum\limits_{i=1}^{N_\mathrm{partons}} \frac{1}{2 \pi v^2} \exp\left[{-}\frac{(\mathbf{x} - \mathbf{x}_i)^2}{2 v^2}\right],
\end{equation}
where $N_\mathrm{partons}$ is the number of partons, $v$ is their width, and $\mathbf{x}_i$ the partons' positions which are sampled from a Gaussian distribution of width $r_\mathrm{parton} = \sqrt{w^2 - v^2}$.
The factor $\gamma_i$ is a Gamma random weight factor as before.
Once the nucleon width, parton number, and parton width are specified, we apply Eq.~\eqref{collision_criteria} to each pair of partons and flag each corresponding nucleon as a participant if one or more of its partons collide.
If any parton in a given nucleon participates, \emph{all} partons in that nucleon contribute to the participant thickness function.
Finally, we numerically tune the cross section parameter $\sigma_{gg}$ which now modulates the parton-parton interaction probability in order to recover the desired p+p inelastic cross section.
Figure~\ref{fig:substructure} shows several examples of the nucleon substructure effect on proton and lead nuclear thickness functions.

\section{Model calibration and results}

The aforementioned substructure extension includes a number of free parameters, e.g.\ the parton number and parton width, which characterize the initial conditions.
These parameters form a complex, multi-dimensional design space which renders manual and brute force optimization methods prohibitively expensive. 
To circumvent this issue, we apply established Bayesian methodology \cite{Higdon:2008cmc} and use Monte Carlo methods to sample the posterior distribution of the model parameters, calibrated to fit p+Pb collision data at $\sqrt{s_\mathrm{NN}} = 5.02$~TeV measured by the ALICE experiment \cite{Abelev:2014mda}.
This procedure, described at length in Ref.~\cite{Bernhard:2016tnd}, is briefly summarized as follows.

\begin{figure}
  \includegraphics{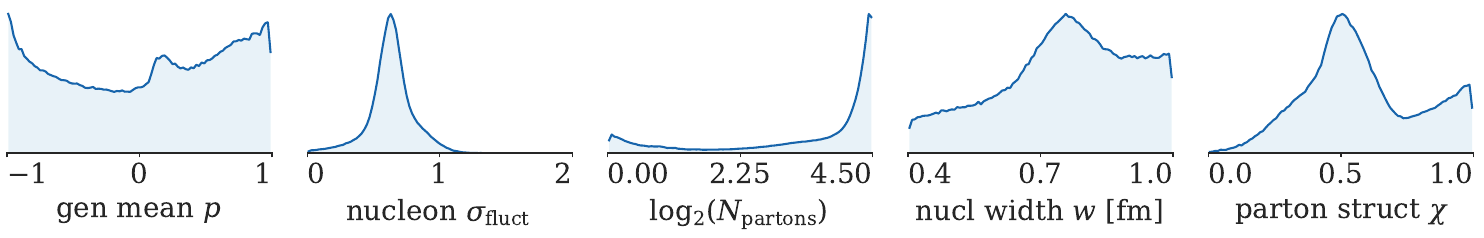}
  \caption{
    \label{fig:posterior} Marginalized posterior distributions for the calibrated model.
    Parameters listed from left to right: generalized mean parameter $p$, nucleon fluctuation standard deviation $\sigma_\mathrm{fluct}$, parton number $\log_2(N_\mathrm{partons})$, nucleon width $w$, and parton structure parameter $\chi$.
    Prior distributions were uniform (flat).
  }
\end{figure}

We first sample $n=240$ unique parameter configurations using a space filling algorithm to construct a scaffolding of the design space, such that each parameter is sampled from a uniform distribution over a finite range.
The parton width cannot be larger than the nucleon width and hence requires special attention.
We reparametrize it using a structure parameter 
\begin{equation}
  v = v_\mathrm{min} + \chi\, (v_\mathrm{max} - v_\mathrm{min}), 
\end{equation}
where the minimum parton width $v_\mathrm{min}=\sqrt{A_\mathrm{min}/N_\mathrm{partons}}$ with $A_\mathrm{min}=0.09 ~\mathrm{fm}^2$ is chosen to always permit a configuration which satisfies the inelastic p+p cross section, and the maximum allowable parton width is capped at $v_\mathrm{max} = w$ the nucleon width.

We then run the \trento\ initial condition model with partonic substructure and calculate the initial entropy density $s$ and eccentricity harmonics $\e_2$ and $\e_3$ for a large number of minimum bias events using the parameter values at each design point.
A Gaussian process emulator is trained to interpolate these quantities as a function of the input parameters.
The trained emulator admits essentially instant predictions for the entropy density  and eccentricity harmonics in a given centrality bin at arbitrary points in parameter space and acts as a fast surrogate for the full physics model.

To make contact between initial state properties and final state observables, we generate a large sample of p+Pb initial condition events using randomly sampled substructure parameters and evolve them through event-by-event hybrid model calculations which couple boost-invariant viscous hydrodynamics to a hadronic afterburner \cite{Shen:2014vra}.
We then measure the yield and flow response functions 
\begin{align}
  \label{response}
  N_\mathrm{ch}(|\eta| < 1) = f(dS/dy), \\
  \label{hydro_response}
  \frac{\vn}{\en} = g\left(\frac{N_\mathrm{ch}(|\eta| < 1)}{A_0}\right),
\end{align}
where $v_n$ is the anisotropic flow coefficient, $\varepsilon_n$ the corresponding eccentricity harmonic, and $A_0$ is the transverse area of the entropy profile, measured by counting the cells above 1\% of the peak entropy density. 
The response functions $f$ and $g$ are finally interpolated with a set of basis splines and used to translate the predictions of the Gaussian process emulator from initial state properties to final state observables. 

Finally, the model is calibrated using the Gaussian process emulator and response functions to perform random walks through the parameter space, weighted by the Bayesian posterior probability that the model predictions describe the ALICE data; we calibrate to both the charged particle yields $N_\mathrm{ch}(|\eta| < 1)$ and flow cumulants $v_n\{2\}$ in several centrality bins \cite{Abelev:2014mda}.
By using a large number of walkers and a large number of steps, we are able to generate a histogram of the walker occupancy density in the multidimensional parameter space.
The resulting posterior distribution is visualized in Fig.~\ref{fig:posterior} which shows projections of the posterior distribution along each model parameter.
We note that the initial condition normalization, which is not shown in Fig.~\ref{fig:posterior}, is fixed to reproduce average particle production across the centrality range.

\begin{figure}
  \centering
  \includegraphics{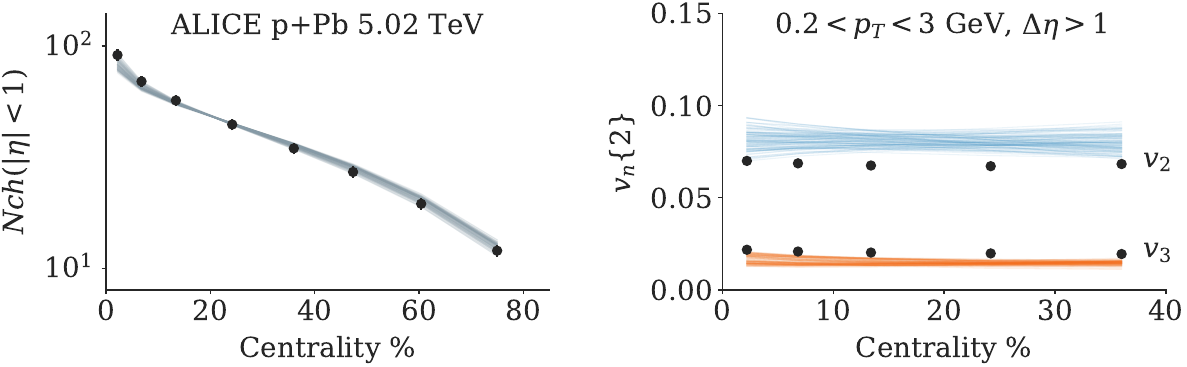}
  \caption{
    \label{fig:observables} Emulator predictions of 100 random samples drawn from posterior distribution compared to experimental data from the ALICE experiment for p+Pb collisions at $\sqrt{s_\mathrm{NN}} = 5.02$~TeV \cite{Abelev:2014mda}.
    Left column: charged particle yield $N_\mathrm{ch}(|\eta| < 1)$; right column: two-particle cumulant $v_n\{2\}$ for $0.2 < p_T < 3.0$~GeV and $\Delta \eta > 1$. ALICE data has been rebinned to construct more uniform centrality intervals.
  }
\end{figure}

Several interesting features emerge from the calibrated posterior.
First, the entropy deposition parameter $p$, which is highly constrained by Pb+Pb data \cite{Bernhard:2016tnd}, is unconstrained by the proton-lead data alone and accepts scaling behavior ranging from a wounded-nucleon model $p=1$ to models with strong saturation effects described by $p < 0$.
Similarly, the nucleon width $w$ is unconstrained by the present analysis, although it does require rather specific values for the nucleon fluctuation standard deviation $\sigma_\mathrm{fluct}$.
We also observe a moderate preference for a large number of partons $N_\mathrm{part} \gtrsim 10$ near the upper end of our design range with a preferred parton width roughly equal to one-half the specified nucleon width.
The performance of the calibrated model is illustrated in Fig.~\ref{fig:observables} which shows emulator predictions plotted against ALICE experimental data \cite{Abelev:2014mda} using parameter values sampled from the Bayesian posterior.
Our calibrated model provides a good description of the charged particle yield and flow cumulants as a function of collision centrality, although we somewhat overshoot the gap of $v_2$ and $v_3$ measured by experiment.
This tension may arise from limitations of the hydrodynamic response scaling in Eq.~\eqref{hydro_response} and merits further investigation.

\section{Summary}
We construct a simple multi-parton description of QGP initial conditions for small collision systems using a variable number of partons and predetermined parton width.
Bayesian parameter estimation is then used to constrain free parameters of the model.
Our results indicate a moderate preference for a parton-like description of the nucleon, although we cannot yet exclude a partonless wounded-nucleon model based on proton-lead data alone.
Natural targets for improvement to the present analysis include an expanded design range, more rigorous emulator training and validation, simultaneous calibration to proton-lead and lead-lead systems, and replacing our response functions with model calculations at each design point.
We leave these refinements to future work.

This work was supported by the U.S.\ DOE Grants No.\ DE-FC52-08NA28752 and DE-FG02-05ER41367 and NSF Grant No.\ NSF-ACI-1550225. CPU time was provided by the Open Science Grid, supported by the DOE and NSF. 

%% References


\bibliographystyle{elsarticle-num}
\bibliography{proceeding}

\begin{thebibliography}{1}
\expandafter\ifx\csname url\endcsname\relax
  \def\url#1{\texttt{#1}}\fi
\expandafter\ifx\csname urlprefix\endcsname\relax\def\urlprefix{URL }\fi
\expandafter\ifx\csname href\endcsname\relax
  \def\href#1#2{#2} \def\path#1{#1}\fi

\bibitem{Higdon:2008cmc}
D.~Higdon, J.~Gattiker, B.~Williams, M.~Rightley, J.Amer.Stat.Assoc. 103~(482)
  (2008) 570.

\bibitem{Moreland:2014oya}
J.~S. Moreland, J.~E. Bernhard, S.~A. Bass, Phys. Rev. C92~(1) (2015) 011901.

\bibitem{Bernhard:2016tnd}
J.~E. Bernhard, J.~S. Moreland, S.~A. Bass, J.~Liu, U.~Heinz, Phys. Rev.
  C94~(2) (2016) 024907.

\bibitem{Schenke:2014zha}
B.~Schenke, R.~Venugopalan, Phys. Rev. Lett. 113 (2014) 102301.

\bibitem{Abelev:2014mda}
B.~B. Abelev, et~al., Phys. Rev. C90~(5) (2014) 054901.

\bibitem{Shen:2014vra}
C.~Shen, Z.~Qiu, H.~Song, J.~Bernhard, S.~Bass, U.~Heinz, Comput. Phys. Commun.
  199 (2016) 61--85.

\end{thebibliography}



%\begin{thebibliography}{00}
%\bibitem{jonah_mtd}
%
%\end{thebibliography}

\end{document}